\documentclass[11pt,preprint]{aastex}
\usepackage{amsmath}
\usepackage{amssymb}
\usepackage{graphicx}  
\newcommand{\bea}{\begin{eqnarray}}  
\newcommand{\eea}{\end{eqnarray}}  
\newcommand{\be}{\begin{equation}}  
\newcommand{\ee}{\end{equation}}  
  
\newcommand{\rt}[1]{{}}

\def\siml{\lower4pt \hbox{$\buildrel < \over \sim$}}
\def\simg{\lower4pt \hbox{$\buildrel > \over \sim$}}
\def\Mesz{M\'esz\'aros~}
\def\eps{\epsilon}
\def\E9{{\eps_9}}
\def\E11{{\eps_{11}}}

\def\P3{P_{-3}}
\def\a6{a_6}
\def\B15{B_{15}}
\def\Z26{Z_{26}}
\def\A56{A_{56}}
\def\L35{L_{35}}
\def\D10{D_{10}}
\def\para{\parallel}
\def\msun{M_\odot}

\def\bec{\begin{center}}
\def\enc{\end{center}}
\def\beq{\begin{equation}}
\def\enq{\end{equation}}
\def\bea{\begin{eqnarray}}
\def\ena{\end{eqnarray}}

\begin{document}

\title{High Energy Neutrinos and Photons from Curvature Pions in Magnetars}
\author{ 
T. Herpay\altaffilmark{1,2},
S. Razzaque\altaffilmark{3,4,5,6}, 
A. Patk\'os\altaffilmark{1,2},
and
P. \Mesz\altaffilmark{3,4,5}
}
\altaffiltext{1} {Institute of Physics, E\"otv\"os University, H-1117 Budapest, Hungary}
\altaffiltext{2} {Research Group for Statistical and Biological Physics of the Hungarian Academy of Sciences, H-1117 Budapest, Hungary}
\altaffiltext{3}{Department of Astronomy and Astrophysics, Pennsylvania State University, University Park, PA 16802, USA}
\altaffiltext{4}{Department of Physics, Pennsylvania State University, University Park, PA 16802, USA}
\altaffiltext{5} {Center for Particle Astrophysics, Pennsylvania State University, University Park, PA 16802, USA}
\altaffiltext{6} {Space Science Division, Code 7653, U.S. Naval Research Laboratory, Washington DC 20375, USA}

\begin{abstract}

We discuss the relevance of the curvature radiation of pions in
strongly magnetized pulsars or magnetars, and their implications for
the production of TeV energy neutrinos detectable by cubic kilometer
scale detectors, as well as high energy photons.

\end{abstract}


\section{Introduction}

\label{sec:intro}

In pulsar and magnetar polar caps one expects the formation of vacuum
or space-charge limited gaps (e.g. Harding and Lai, 2006), in which an
electric field component parallel to the magnetic field accelerates
particles to relativistic velocities.  The particles move away from
the caps along the magnetic field lines, producing photons by
curvature radiation, which at some height $h$ produce an $e^\pm$ pair
front that determines the upper extent of the gap (e.g. Ruderman and
Sutherland, 1975, Harding and Muslimov, 2002). The caps with
$\Omega\cdot B<0$ can accelerate positively charged protons or ions,
whose interaction with X-rays from the polar cap will lead to very
high energy (VHE, $E\simg 10^{12}$ eV) neutrinos through photomeson
interactions, $p\gamma \to \pi^+ \to \mu^+\nu_\mu \to e^+\nu_e{\bar
\nu}_\mu$ (Zhang et al, 2003).  Here we explore the consequences for
neutrino production of a different mechanism, the curvature radiation
of pions by protons.

Relativistic protons interacting with a strong magnetic field can
produce pions, a quantum treatment of this process being given by
Zharkov (1964). Ginzburg and Zharkov (1965) used a semi-classical
approach to emphasize the analogy between this pion radiation process
and the usual synchrotron radiation of photons by protons.  This
semi-classical method was further explored for $\pi^0$ and $\rho$
mesons by Tokuhisa and Kajino (1999) in the context of pulsars. The
pion radiation mechanism is characterized by a parameter
$\chi=\gamma^2 (B/B_Q)$ similar to that used for electron processes
but with the proton mass replacing the electron mass, with $B_Q=
m_p^2c^3/e\hbar=1.5 \times 10^{20}$ G. They showed that the
synchrotron-like $\pi^0$ radiation process becomes competitive with
(proton) photon synchrotron radiation when the above parameter $\chi
\simg 0.1$. The related synchrotron-like $\pi^+$ radiation by protons
in a magnetic field requires, however, a different semi-classical
treatment, since the protons become neutrons in the process, and this
was calculated by Herpay and Patk\'os (2008).

 In the astrophysical literature, particle acceleration in pulsars or
magnetars is considered both near the polar caps (``inner" or polar
gap models), and further out near or beyond the light cylinder (``outer" 
gap or wind models); e.g. see review by Harding and Lai (2006).
Examples of the former are e.g. Blasi, Epstein and Olinto (2000), Arons 
(2003), etc.; while examples of the latter are, e.g. Sutherland (1979),
Harding and Muslimov (2001, 2002, 2005), Thompson (2008), etc.. Here, 
we consider specifically the case of protons or ions accelerated in 
the inner regions, just above the polar caps of magnetars. 
In the inner acceleration region discussed here, the accelerated 
particles are expected to radiate away their transverse energy so
efficiently that they are effectively constrained to the ground Landau
level as they move along the curved field lines. The principal photon
radiation loss is then curvature radiation (Ruderman and Sutherland,
1975; Harding and Lai, 2006). In this situation, the synchrotron-like
pion radiation (which depends on the proton transverse energy) will be
similarly suppressed. However, as for photon curvature radiation, the
motion along the curved field lines produces an acceleration
equivalent to a Larmor motion in a fictitious magnetic field, which
will lead to curvature pion radiation (Berezinsky, Dolgov and
Kachelriess, 1996). Unlike previous authors, in this paper we
concentrate on $\pi^\pm$ production by the curvature-like mechanism of
both protons and heavy nuclei, occurring in the immediate vicinity of
the polar caps, and we explore its consequences for the production of
VHE neutrinos, as well as their detectability by cubic kilometer
Cherenkov detectors such as IceCube and KM3NeT. We compare the
resulting neutrino fluxes to those from the photomeson process, and
discuss the associated cosmic ray as well as GeV-TeV photon
production.

In \S \ref{sec:model} we discuss an approximate acceleration and radiation 
loss model, including both photon and pion radiation losses.
Simple time-scale arguments are presented to estimate the maximum particle
energies possible and the ratio of photon to pion losses. In \S \ref{sec:num}
we consider in more detail the equations of motion in self-consistent
polar cap  magnetic field models, and calculate numerically the acceleration
and energy loss rates in a general manner. These results are in good
agreement with the approximate estimates of the previous section. In \S 3 and
\S 4 we estimate the resulting neutrino luminosity from curvature
pion decays and from photo-meson effects. In \S \ref{sec:disc} we
discuss the possibility of detecting these effects in kilometer scale neutrino
detectors and at GeV photon energies.

\section{Ion Acceleration and Energy Losses}

\label{sec:model}

The maximum potential drop in a magnetar of radius $a=10^6\a6$ cm,
rotation frequency $\Omega=2\pi/P$ (where $P=10^{-3}\P3$ s is the
period) and surface magnetic field at the pole $B=10^{15}\B15$ G is
$\Phi=(\Omega^2 B a^3/2 c^2)\simeq (6.6\times 10^{21}~\hbox{V}) \B15
\a6^3 \P3^{-2}$. However, the self-consistent gap structure of
magnetars is not well known, e.g. for the ${\vec{\Omega}\cdot \vec{B}}
<0$ caps where proton acceleration can occur. Vacuum gaps may be
expected (Medin and Lai, 2008) under some assumptions about the
surface cohesive energy and thermal structure. On the other hand, the
uncertainties and approximations involved leave open the possibility
of space-charge limited flow (SCLF) gaps, and in this paper we assume
that the latter situation applies.  Ions may be lifted from the
magnetar surface and accelerated by an unsaturated electric field
component parallel to the magnetic field lines for SCLF gaps
(e.g. Harding and Muslimov 2002) up to a height similar to the polar
cap radius $\sim r_{\rm pc}=(2\pi a^3/cP)^{1/2} \approx 4.6\times 10^5
\a6^{3/2}\P3^{-1/2}$~cm.  Beyond a height $h\sim r_{\rm pc}$, ions may
further be accelerated by an electric field in the saturated regime,
given by $E_\para \simeq 5\times 10^{12} \B15 \P3^{-2}$ in
c.g.s. units (see, e.g., eq.~1 in Harding and Muslimov 2002).
For heights much smaller than the stellar radius, $h \ll a$ for which
the saturated field is constant, the energy $\eps$ reached by the
particle is $\eps= Ze E_\para h$, during an acceleration time
\beq 
t_{a} = \epsilon / c Z e E_\para \simeq 2.2 \times 10^{-8} \eps_9\B15^{-1} \P3^{2} Z^{-1}~\hbox{s},\label{eq:tap} 
\enq 
where $\eps_9=(\eps/10^9 \hbox{eV})$.  Charged particles accelerated
from the polar cap move following closely the field lines, the main
photon radiation loss being curvature radiation, whose power is
${\cal P}_c=(2 e^2 c \gamma^4 / 3 R_c^2)$.  Curvature radiation generally 
dominates over (transverse) synchrotron radiation at low heights
in the polar caps. 
For the limiting open field line of a neutron star rotating with period 
$P=10^{-3}P_{-3}$ s defining a light cylinder $R_L=cP/2\pi$ the
curvature radius is
\beq
R_c=(4/3)(a R_L)^{1/2}=3\times 10^6 \P3^{1/2} \a6^{1/2} ~\hbox{cm}.\label{eq:Rcurv}
\enq 
and the photon curvature radiation loss time for protons of energy
$\eps$ is
\beq 
t_{c} = (3 \eps R_c^2/ 2 e^2 c \gamma^4) = 2.4\times 10^{-9}\eps_9^{-3} \P3 \a6 ~\hbox{s}.\label{eq:tcp}
\enq
Another energy loss mechanism for protons in a strong magnetic field
is the synchrotron-like pion radiation losses (Ginzburg and Zharkov, 1964).
In terms of the parameter $\chi$ defined as
\beq 
\chi=\gamma(B/B_Q),\label{eq:chi}
\enq
where $B_Q=m_p^2c^3/e\hbar =1.5 \times 10^{20}$ G, a semi-classical calculation in the $\chi\gg 1$ asymptotic limit gives the energy loss rate ${\cal P}_{p,\pi^0}=dE_p/dt$ of a proton due to $\pi^0$ emission as
\beq 
{\cal P}_{\pi^0}(\chi\gg 1) \simeq (g^2 /6) m_p^2 c^3 \hbar^{-2},\label{eq:Ppi0gg1}
\enq
which in quantum mechanical calculation acquires a further $\chi^{2/3}$ 
factor (Ginzburg and Zharkov, 1964). Here $g^2/\hbar c \simeq 14$ is the 
strong coupling constant. The corresponding low $\chi$ limit is
\beq 
{\cal P}_{\pi^0}(\chi\ll 1) = (g^2 /\sqrt{3}) m_\pi m_p c^3 \hbar^{-2}\chi \exp[-(\sqrt{3}/\chi)(m_\pi/m_p)] \label{eq:Ppi0ll1}
\enq
(Ginzburg and Zharkov, 1964; Tokuhisa and Kajino, 1999).  For
$\chi\simg 0.1$ these losses can exceed those from photon synchrotron
radiation.  In the related process of $\pi^+$ meson radiation, the
proton becomes a neutron during the emission process, and a different
semi-classical calculation method is needed, as done by Herpay and
Patk\'os (2008). A quantum calculation of this process is that of
Zharkov (1964). In the limit of $\chi\ll 1$ which is of interest for
us here, the proton energy loss rate due to $\pi^+$ radiation is
\bea
{\cal P}_{\pi^+} & = & \left({g^2 \over \hbar c}\right)\left({m_p^2 c^4\over \hbar}\right) \left({ 1\over \sqrt{24} }\right) \left[{ (m_p/m_\pi)^3(2\sqrt{2}-m_\pi/m_p) \over { \left[1+\sqrt{2}(m_p/m_\pi) \right]^4 } }\right] \chi \exp \left[ -{\sqrt{3}\over\chi} \left({m_\pi \over m_p}\right)^2 \right]\nonumber \\ &\simeq & 4.15 \times 10^{20} \chi \exp [-0.039/\chi ]~\hbox{erg s$^{-1}$}.\label{eq:Ppi+ll1}
\ena
However, for the same reason that photon curvature radiation dominates
over photon synchrotron radiation, the pion radiation due to the proton
transverse momentum will be dominated by pion radiation due to the
longitudinal component along the curved field lines (Berezinsky et al, 1996).
 We define an equivalent `curvature" $B_\perp$ field perpendicular to
the particle trajectory, which would give a radius of curvature $R_c$
equal to the gyro-radius of a proton of energy $\eps$ and Lorentz
factor $\gamma$,
\beq 
B_\perp= \eps / e R_c = 1.1\times 10^9 \eps_9 \P3^{-1/2} \a6^{-1/2} ~\hbox{G}.\label{eq:Bperp}
\enq
The corresponding `curvature" $\chi$ parameter is 
\beq 
\chi_\perp=\gamma(B_\perp /B_Q) \equiv \gamma^2 (\hbar/ R_c m_p c)   \simeq 7.84\times 10^{-3} \eps_9^2 \P3^{-1/2} \a6^{-1/2} ~,\label{eq:chiperp}
\enq
and in terms of this parameter the photon curvature power is
\beq
{\cal P}_{\gamma,c} = (2 e^2 c \gamma^4 / 3 R_c^2) \equiv (2 e^2/3\hbar c)(m_p^2 c^4/\hbar) \chi^2 ~.\label{eq:Pcurvchi}
\enq
From now on we will refer to $\chi\equiv \chi_\perp$, and in our case
we will be interested mainly in the $\chi\ll 1$ limit.
Within the range  of this inequality it is reasonable to assume that 
the small momentum transfer regime applies and provides an adequate 
approximation to the interaction rate. This is the region where the simple 
effective models leading to equations (\ref{eq:Ppi0ll1}) and (\ref{eq:Ppi+ll1}) 
are valid. This intuitive description is based on the assumption that the
transverse momentum of the emitted pions is of the same order of
magnitude as that of the proton moving along the curved trajectory in
the strong  magnetic field (Berezinsky et al, 1996). An accurate
discussion of this issue is still an open question.
 In this $\chi \ll 1$ limit, 
the cooling time against pion curvature radiation of a particle of
energy $\eps$, $t_{\pi,c}= \eps /{\cal P}_{\pi,c}$,  using eq. 
(\ref{eq:Ppi+ll1}) with eq.(\ref{eq:chiperp}), is
\beq 
t_{\pi,c}\sim \begin{cases}8.35\times 10^{-14} \eps_9^{-1} \P3^{1/2}\a6^{1/2} 
  \exp[31.76 \eps_9^{-2} \P3^{1/2} \a6^{1/2}]~\hbox{s} &{\rm for}~ \pi^0 ; \cr 
4.27\times 10^{-13} \eps_9^{-1} \P3^{1/2}\a6^{1/2} \exp[4.846 \eps_9^{-2} \P3^{1/2} 
  \a6^{1/2}]~\hbox{s} &{\rm for}~ \pi^+ \end{cases}
\label{eq:tpi}
\enq
where $\eps_9=(\eps/10^9\hbox{GeV})$ is the particle (longitudinal) energy. 

For the case of heavy ions $(A,Z)$, e.g. Fe (56,26), for an ion of energy $\eps$ 
the ion Lorenzt factor is $\gamma_A=\eps/(Am_p c^2)$, and each constituent proton 
moves with $\gamma_A$. We can use the previous equations for the photon curvature 
losses ${\cal P}_{\gamma,c}$, and in principle also the curvature pion losses
${\cal P}_{\pi,c}$ of the individual constituent protons moving with $\gamma_A$, 
and find the corresponding losses for the ion as ${\cal P}_{\gamma,c,A} =Z 
{\cal P}_{\gamma,c}(\gamma_A)$.  The proportionality factor $Z$ assumes incoherent 
radiation by the constituent nucleons, since at high energies the dominant 
wavelength becomes smaller than the average separation of the constituents. 
Thus ${\cal P}_{\gamma,c,A}= Z {\cal P}_{\gamma,c} = Z (2/3) (e^2/\hbar c) 
(m_p^2 c^4 /\hbar)\chi^2$.  For an ion energy $\eps_A=10^{11}\eps_{11}$ GeV, the 
acceleration and incoherent photon curvature times are
\bea 
t_{a,A} &=& 8.6\times 10^{-8} \eps_{11} \B15^{-1} \P3 \Z26^{-1} ~\hbox{s}, \\ 
t_{\gamma,c,A} &=& 1.07 \times 10^{-9} \eps_{11}^{-3} \P3\a6\A56^4 \Z26^{-1}~\hbox{s}, 
\label{eq:tZ}
\ena  
where we used the same curvature radius $R_c$ of equation (\ref{eq:Rcurv}) 
as for protons, since the nuclei move along the same field lines. 
The fictitious `curvature" magnetic field $B_\perp$ and $\chi_\perp$ for ions 
is the same eq.~(\ref{eq:Bperp}) and eq.~(\ref{eq:chiperp}) as for protons,
except for using $\gamma_A$,
\beq 
\chi_A=\gamma_A (B_\perp/B_Q)= 2.62\times 10^{-2} \eps_{11}^2 A_{56}^{-2} (a_6 P_{-3})^{-1/2} ~.\label{eq:chiZ}
\enq
 If the ions did not fragment before reaching the appropriate energy
threshold, the pion curvature radiation of heavy ions $(A,Z)$ moving along 
the field lines would, in principle, be due to the individual nucleons in the 
nucleus moving with Lorentz factor $\gamma_A$ and with the $\chi$ parameter  
of eq. (\ref{eq:chiZ}). The pion emission power would then be 
${\cal P}_{\pi,c,A}=A {\cal P}_{\pi,c}$ for the above $\chi_A$, the protons 
in the nucleus radiating $\pi^+$ and the neutrons radiating $\pi^-$.
 However, as discussed in \S 3.1 and \S 3.2 , the interaction with the curved 
magnetic field and the radiation recoil of individual nucleons 
is found to exceed the binding energy of the nucleus, leading to its
fragmentation, before reaching the energy at which the nucleus could
radiate pions. The individual resulting protons would, however, continue 
to be accelerated and would radiate individually, as described before.

For protons, it is seen from Fig. \ref{fig:tp} that the intersection
of the $\pi^0$ energy loss timescale with the acceleration time occurs
generally at higher energies than the intersection of the $\pi^+$
losses and the acceleration. The shorter $\pi^+$ time is mainly due to
the lower energy cutoff in the radiation power ${\cal P}_{\pi^+}$
caused by an $(m_{\pi^+}/m_p)^2$ factor in the exponent, versus an
$m_{\pi^0}/m_p$ in the exponent of the $\pi^0$ power. Thus, the proton
energy is determined by a balance between acceleration energy gains
and a combination of the $\pi^+$ radiation and photon curvature
losses.  Assuming that the particles can reach the corresponding
height $h$ (which is plausible in the saturated field regime 
$r_{\rm pc}\ll h\ll a$), the maximum particle energy for a given value 
of $B$ and $P$ is approximately given by the intersection of the
appropriate acceleration time with either the $\pi^+$ or the curvature
loss time curve, whichever occurs at the lowest energy. Thus, from
Fig. \ref{fig:tp}, for $B=10^{15}$ G and $P=10^{-3}$ s the curvature
pion losses are beginning to get competitive with curvature photon
losses, with $t_{\pi^+}/t_c\sim 0.2$ at $\eps_{max}\sim 0.65 \times
10^9$ GeV. The pion losses dominate for $\B15\P3^{-2}\simg 2$.
However for lower fields or longer periods $\B15\P3^{-2} \siml 1$, it
is the photon curvature losses that determine $\eps_{max}$.


\section{Trajectories, Acceleration and Energy Losses from the Solution of the Proton/Ion Equations of Motion}

\label{sec:num}

In the previous section, we determined the maximum energy of the
protons/ions by comparing the characteristic timescales of different
radiations with that of the acceleration calculated without taking
into account the back-reaction of the radiation. Moreover, in
estimating the characteristic times we made the further assumption that
the perpendicular momentum of the accelerated particle to the magnetic
field lines is much smaller than the parallel component.  The
simplicity of this physical picture greatly helps our intuition, still
its fundation needs some elaboration. Therefore it is interesting to
compute the trajectory and the saturation energy of the charged
particles directly from the full equation of motion which also takes
into account the backreaction of the radiation.  In this section we
discuss in detail the proton/ion trajectory through the acceleration
region near the polar cap.

\subsection{Equations}

In a given electromagnetic field, the relativistic equation of motion
of a proton is

\be
\dot{\vec{\gamma}}=\frac{e}{m_pc}\left(\vec{E}+\frac{1}{\sqrt{1+\vec{\gamma}^2}} \vec{\gamma}\times \vec{B}\right) - \frac{{\cal P}}{m_pc^2}\frac{\sqrt{\vec{\gamma}^2+1}}{|\vec{\gamma}|}\frac{\vec{\gamma}} {|\vec{\gamma}|},\qquad \vec{\gamma}=\frac{ {\vec{p}}}{{m_pc}},\label{e:eom}
\ee
where we added on the right hand side the last term to the usual
Lorentz force to take into account the energy loss of the particle
(${\cal P}=\mathrm{d}E/\mathrm{d}t$) due to radiation processes
(photon, $\pi^0$, $\pi^+$). The direction of this back reaction force
is chosen opposite to its momentum, which is true for large
$|\vec{\gamma}|$ (a similar description of the back reaction of the
radiation was proposed in Takata et al (2008) and Harding, Usov and
Muslimov (2005)). Here, $\cal P$ results from the curvature of the trajectory. It is not
split up into separate synchrotron and curvature (along $\vec{B}$)
radiation parts corresponding to transverse and parallel momentum
components, respectively. This approach enables us a more complete
description, because the presence of an accelerating electric field
continously modifies the curvature even if $\vec{E} || \vec{B}$. Therefore in eq.~(\ref{e:eom}) $\cal P$ depends on the total curvature radius ($R_c$), which defines a curvature parameter $\chi$ (Berezinsky et al, 1995),
\be
\chi=\frac{\vec{\gamma}^2}{R_c}\frac{\hslash}{m_p c},
\ee
where $R_c$ can be expressed with $\vec{\gamma}$ and $\dot{\vec{\gamma}}$,
\be
\frac{1}{R_c}=\frac{1}{c} \frac{\sqrt{1+\vec{\gamma}^2}}{|\vec{\gamma}|^3} \left|\dot{\vec{\gamma}}\times\vec{\gamma}\right|=\frac{1}{c} \frac{\sqrt{1+\vec{\gamma}^2}} {|\vec{\gamma}|^3}\sqrt{\vec{\gamma}^2{\dot{\vec{\gamma}}}\!\left.^{2}\right. -(\vec{\gamma} \dot{\vec{\gamma}})^2}.
\ee
Here {\cal P} is the sum of the power of $\gamma$ and $\pi$ radiations given in section 2:
\be
{\cal P}={\cal P}_c+{\cal P}_{\pi^0}+{\cal P}_{\pi^+}\equiv {\cal P}\left(|\vec{\gamma}|,\chi(\vec{\gamma},\dot{\vec{\gamma}})\right), \label{e:P}
\ee
The formulae are applicable only for $\chi<m_\pi/m_p$. It can be seen
from the above expressions that $\dot{\vec{\gamma}}$ appears on the
right hand side of the equation of motion (\ref{e:eom}) in a very
complicated way. However if the ratio of the power of the energy loss
to the energy is small, one can solve the equation iteratively,
e.g. one substitutes $\dot{\vec{\gamma}}$ into $\chi$ with its value
calculated from the equation of motion with only the Lorentz term (without $\cal P$).

In this approximation one considers six coupled first order
differential equations for $\vec{r}$ and $\vec{\gamma}$ wherein time
derivatives appear only on the left hand sides
\be
\dot{\vec{r}}=\frac{c\vec{\gamma}}{\sqrt{1+{\vec{\gamma}}^2}}\,,\quad \dot{\vec{\gamma}}=\frac{e}{m_pc}\left(\vec{E}(\vec{r}) + \frac{1}{\sqrt{1+\vec{\gamma}^2}} \vec{\gamma}\times\vec{B}(\vec{r})\right)-\frac{{\cal P}(|\vec{\gamma}|,\bar{\chi}(\vec{\gamma}, \vec{r}))}{m_pc^2}\frac{\sqrt{\vec{\gamma}^2+1}}{|\vec{\gamma}|}\frac{\vec{\gamma}}
{|\vec{\gamma}|}, \label{e:final}
\ee
where $\bar{\chi}$ is calculated from (\ref{e:eom}) for ${\cal P}=0$,
\be
\bar{\chi}=\frac{e\hslash}{m_p^3 c^2} \frac{1}{|\vec{\gamma}|}\left|\vec{\gamma}\times \left(\sqrt{1+\vec{\gamma}^2}\,\vec{E}(\vec{r})+\vec{\gamma}\times \vec{B}(\vec{r})\right)\right|.
\ee
One can easily check that this approximation gives back the usual formulas for the intensity of the synchrotron or curvature radiation as long as $\vec{E}=0$ or $\vec{B}=0$, respectively. Note that equations in (\ref{e:final}) would be further simplified if $\vec{E}\,||\,\vec{B}$ (see Takata et al, 2008) using appropriate local coordinates to treat the motion parallel and perpendicular to the magnetic field lines. In the general case
($\vec{E}\times\vec{B}\ne 0$) one cannot avoid to calculate all three independent components of $\vec{\gamma}$. Then the use of a coordinate system with one axis fixed along the magnetic field lines does not bring any extra advantage over the use of a unique coordinate system for the whole motion.

Unfortunately the exact expression of the electromagnetic field to be used in (\ref{e:eom}) is not known fully since the field-charge, the motion and emission of any charged particle modifies the accelerating field. The full selfconsistent description of the inner gap is very complicated as was already shortly discussed in section 2.  In our actual calculations we considered a dipole field for the magnetic part. In addition a so-called unsaturated expression was used for the electric field close to the surface of the magnetar and saturated
field form is adjusted to it far away from the surface. The explicit expressions of the electric field strength in the two regimes are given by Harding and Muslimov (2001, 2002). These electric fields take into account the influence of the accelerated electrons/positrons and their pair production, which may be the most important effect for the formation of a self consistent electric field. The corresponding unsaturated and saturated potentials can be found in equations (20) and (24) of Harding and Muslimov (2001).

In the course of the numerical solutions of (\ref{e:final}) we switched from the unsaturated to the saturated field at the altitude where the component of the unsaturated electric field parallel to the magnetic field equals to the corresponding component of the saturated field.

By consulting the literature it appears, that our choice of the explicit form for the electric field is somewhat arbitrary, many other expressions having been proposed earlier (see e.g. Goldreich and Julian, 1969; Sutherland, 1979; Medin and Lai, 2008 and their references). We checked that the proton/ion saturation energy reflecting a balance between the acceleration and back reaction, and also the power of the different radiations are of the
same order of magnitude when a simple electric quadrupole field is considered instead of the combination of the unsaturated and saturated fields as described above.

The equations discussed above are valid only for protons. In case of accelerated ions, the relevant equation of motion can be arrived at by the replacements $e\longrightarrow Ze$ and $m\longrightarrow A m$ in (\ref{e:eom}).  We study in this section the case  when the radiation intensity by the ions is the incoherent sum of  individual nucleonic contributions provided that both photon and   pion emission processes of the constituents become incoherent when  the ions attain very large energies.  Then one has to make the  following modifications in the expression of $\cal P$ in (\ref{e:P}),
\begin{itemize}
\item[$\bullet$] ${\cal P}_\gamma$ $\longrightarrow$ $Z {\cal  P}_\gamma$ because only protons can emit photons;
\item[$\bullet$] ${\cal P}_{\pi^{0}}$ $\longrightarrow$ $A{\cal P}_{\pi^{0}}$ because both proton and neutron emits $\pi^0$ with the  same probability and neutrons move together with the protons;
\item[$\bullet$] ${\cal P}_{\pi^{+}}$ $\longrightarrow$ $A{\cal P}_{\pi^{+}}$ due to the radiation of $\pi^+$ by protons and the radiation of $\pi^{-}$ by neutrons have same probability (from the point of view of neutrino radiation by charged pions we do not distinguish $\pi^-$ and $\pi^+$).
\end{itemize}

The equation of motion for heavy ions as detailed above is valid as long as the
ion doesn't fragment. Fragmentation can occur as a result of the radiation 
recoil of the constituent nucleons. The numerical solution of the  equation of 
motion  shows that the transverse momentum transfer from the radiation $k_\perp 
\approx \chi^{1/3} m$  exceeds the nucleon's binding energy ($\approx 10$\,MeV 
for Fe ions) before the pion radiation occurs (see the right hand side
of Figs.~\ref{f:ion}).  This  means 
that the main part of the pion radiation arises from the acceleration of the 
individual proton constituents of the dissociated nucleus, as described by equation 
(\ref{e:eom}). Therefore,  one doesn't need to deal with the details of ion 
fragmentation from  the point of view of pion radiation, which would
be given in this treatment by an incoherent sum of the radiation from 
the individual nucleon fragments.
 
In the next subsection we shall present the numerical solution of the equation of 
motion for protons and the corresponding intensity of each type of radiations 
along the trajectory.

\subsection{Calculations and results}

One of the main goals of this calculation is to decide if one can find such values for the magnetic induction $B$ and the rotation frequency $P$ of the neutron star for which the power of charged pion radiation becomes comparable to the photon radiation. Therefore, based on the comparison of the timescales as presented in section 2, we focus our study on $(B,P)$ pairs for which the intersections of the pion and photon cooling time curves with the acceleration time curve happen approximately at the same time.  Based on the findings of the previous
section (see Fig.~\ref{fig:tp}), we examined proton acceleration the following realistic magnetar configurations: ($B_{15},\,P_{-3}$), ($B_{15},\,P_{-2}$), ($B_{14},\,P_{-3}$).

\begin{figure}[t]
\includegraphics[width=0.5\textwidth]{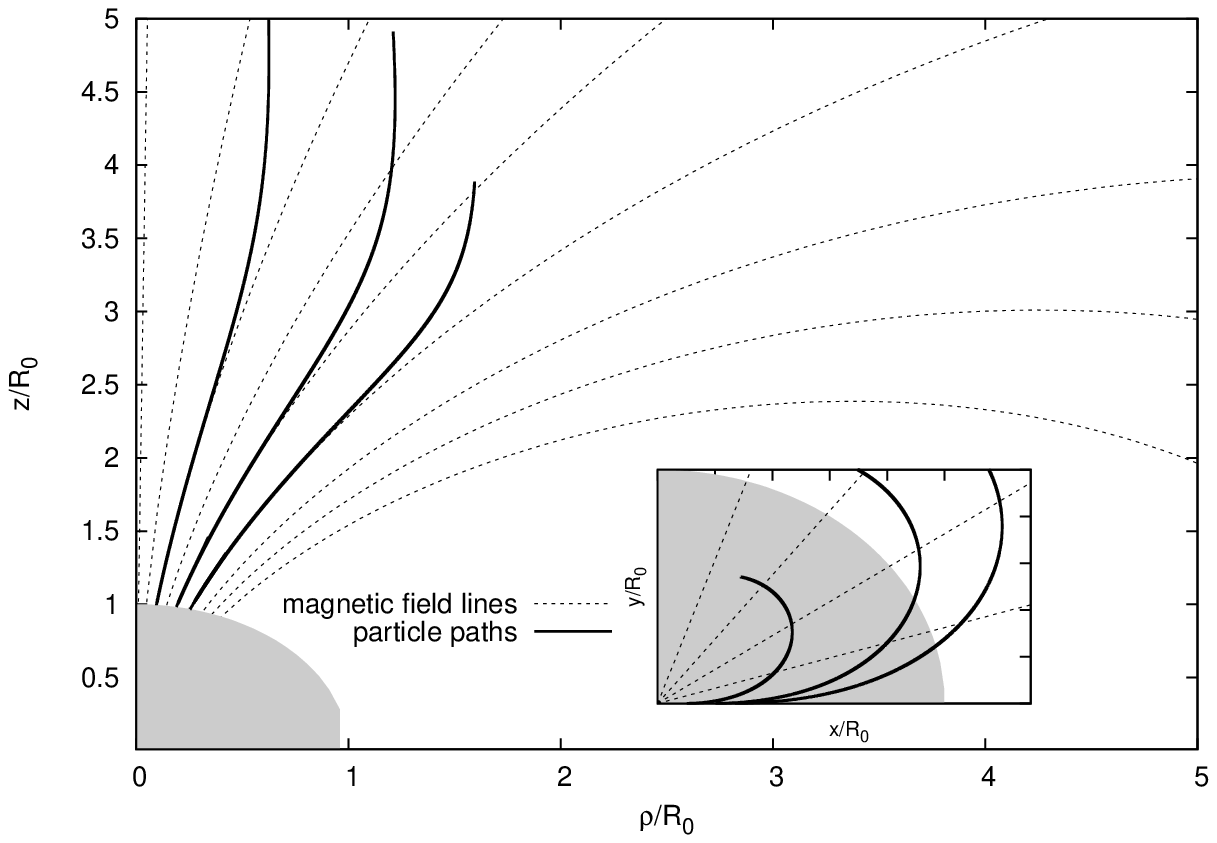}\hfill \includegraphics[width=0.5\textwidth]{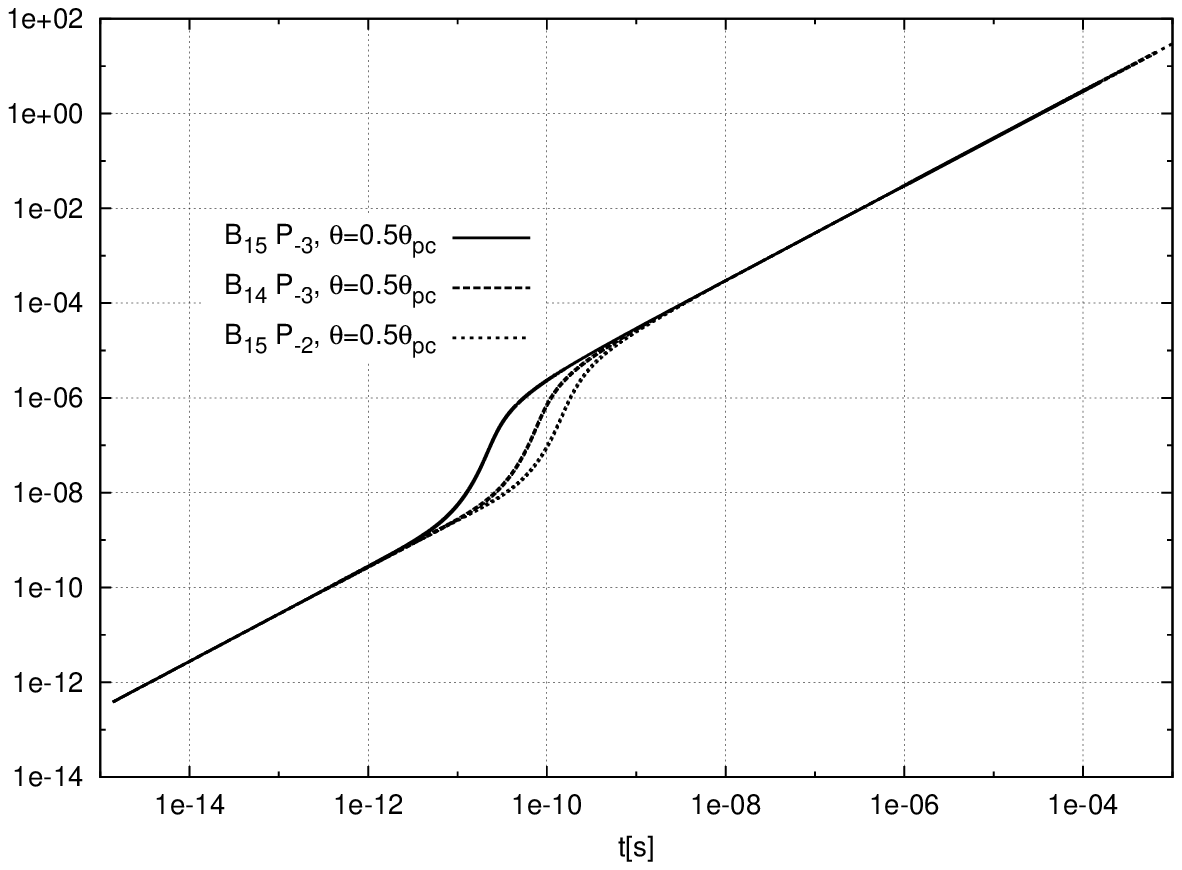}
{\caption{ On the left hand side: the proton's trajectory for initial angles
relative the rotation axis $\theta/\theta_{pc}=0.3,\,0.4,\,0.6$ and
$B_{15}$,\,$P_{-3}$ ($\theta_{pc}=27.25^\circ$). $\theta_{pc}$ is the
angular size of the polar cap. On the right hand side: the time
dependence of the relative altitude of protons above the surface of
the neutron star for different values of $B$ and $P$. \label{f:path}  }}
\end{figure}

The path of a proton is shown for ($B_{15},\,P_{-3}$) and different initial angles $(\Theta)$ measured relative to axis of the magnetic dipole on the left hand side of Fig.~\ref{f:path}. It can be seen that the proton moves initially nearly along the magnetic field lines when the acceleration process begins, then shoves off the
field lines at about $z=2$--$3\,R_0$ and finally moves on a helical trajectory along the z-axis far away from the surface.



\begin{figure}[!b]
\begin{center}
\includegraphics[width=0.83\textwidth]{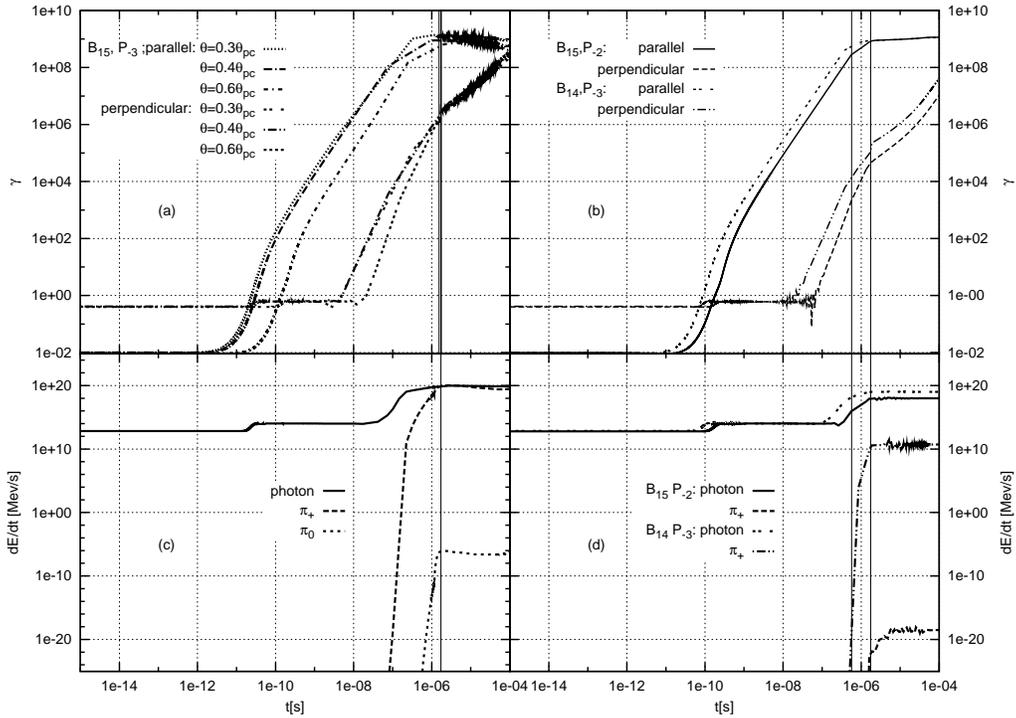}
\end{center}
{\caption{  Acceleration of protons. The parallel and perpendicular momenta for
$B_{15}$,\,$P_{-3}$ (a), and for $B_{15}$,\,$P_{-2}$
($\theta_{pc}=8.33^\circ$) and $B_{14}$,\,$P_{-3}$
($\theta_{pc}=27.25^\circ$) where $\theta=0.5\theta_{pc}$ (b).  The
intensity of each type of radiation for $B_{15}$,\,$P_{-3}$ (c), for
$B_{14}$,\,$P_{-3}$ and $B_{15}$,\,$P_{-2}$ (d) where
$\theta=0.5\theta_{pc}$. The vertical lines indicate the switch
between the unsaturated and saturated fields. \label{f:proton} }}

\end{figure}

On the top of the left hand side of Fig.~\ref{f:proton}, one can see that the maximum proton energy is almost independent of the initial angle when ($B_{15},\,P_{-3}$).  This is also true in case of other values
of $B$ and $P$ and also valid for the intensity maximum of the
radiations as well. Therefore we only plotted the time dependence of
the momenta and the power of radiations corresponding to
$\theta=0.5\theta_{pc}$ (except the top left of Fig.~\ref{f:proton}).

\begin{figure}
\begin{center}
\includegraphics[width=0.495\textwidth]{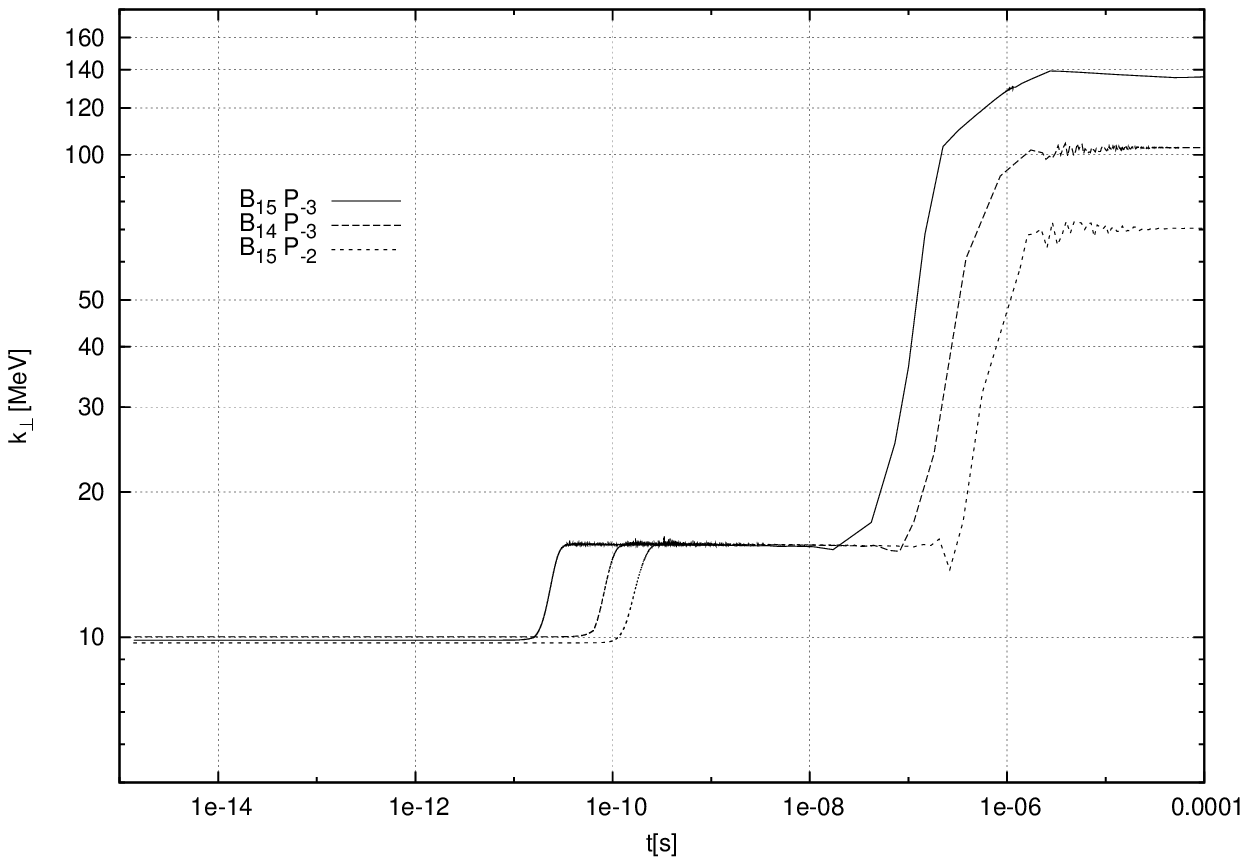}\hfill \includegraphics[width=0.495\textwidth]{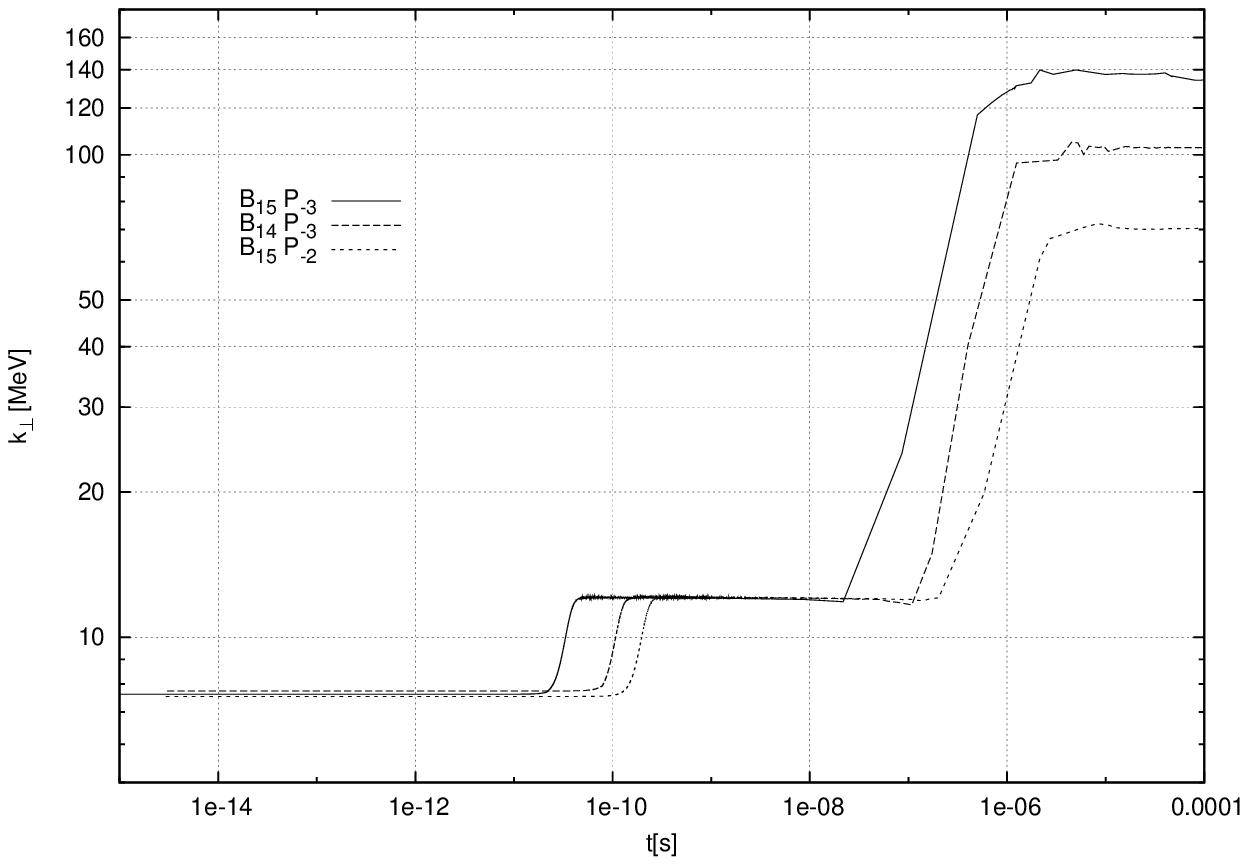}
\caption{Transverse momentum transfer the radiations from protons (left hand side)
  and constituent protons of Fe ions (right hand side) along the path. The photon radiation dominates in case of ($B_{15},\,P_{-2}$)
  and ($B_{14},\,P_{-3}$), and in case of ($B_{15},\,P_{-3}$) up to
  $t\approx 10^{-6}$s. \label{f:ion}} \vspace*{-3cm}
\end{center}
\end{figure}

It's interesting to observe time lags in the growth of both momentum
components parallel/perpendicular ($\gamma_\para/\gamma_\perp$) to the field lines as can be seen on
the top of
Fig.~\ref{f:proton} and on the right hand side of
Fig.~\ref{f:path}, in which the time dependence of the relative
altitude ($(r-R_0)/R_0$) of protons is shown. In this time region, the proton path is closer to circular
($\gamma_\perp \gg \gamma_\para$) and remains roughly constant because
synchrotron radiation loss keeps pace with the transverse energy
gain due to acceleration along the field. Then, the path changes to being more elongated due to the
increasing  electric
field ($|\vec{E}_{unsat}|\sim (r-R_0)/R_0 $), resulting in a change of
the effective radius of curvature. This change of the path is 
indicated by a jump of the value of $\chi=\gamma/R_c$ (at $\approx 10^{-10}$\,s) which can  be
seen in  Fig.~\ref{f:ion}, where we plotted the transverse  momentum
carried by the
transfers of radiations ($k_\perp\approx m \chi^{1/3}$). Between $10^{-10}-10^{-7}$\,s  $\gamma_\perp$ is approximately
constant and the effect from the proton energy gain is practically
compensated by a decrease in the effective curvature radius, so the
radiation rate is approximately constant (see the bottom parts of
Fig.~\ref{f:proton} and Fig.~\ref{f:ion}). The increase in radiation rate
after $10^{-7}$\,s is due to the effective radius of curvature
starting to decrease more slowly than the proton gains energy,
until the saturated regime is reached where energy gains balance
energy losses.





From the bottom parts of Fig.~\ref{f:proton}, one can see that the $\pi_0$ radiation is irrelevant,
moreover the charged pion radiation intensity becomes comparable to
that of the photon radiation only in case of ($B_{15}$,\,$P_{-3}$). 

 In Fig.~\ref{f:ion}, one can see that the transverse momentum transfer of
    the radiation is small comparing to the total energy both
    for protons and  for constituent protons of Fe ions. It supports
    the selfconsistency of our softness assumption for the  radiation process implicit in our choice of the
    effective pion--nucleon interacton. However, $k_\perp$ is large enough
    to fragment the  ions at the beginning of the trajectory as we noticed in the previous sections.

The results of the numerical solution tell us that the protons
can reach energies up to $\approx 10^9$/$10^{10}$\,GeV in the examined
range of B and P. This is in good agreement with the results of the
previous section where we compared the cooling time curves to the
acceleration time curve. This conclusion is apparently not sensitive
to the fact that the value of the momentum perpendicular to the
magnetic field lines is almost the same order of magnitude as the
parallel component due to the essential presence of perpendicular
components of the electric field. We also saw that for
($B_{15},\,P_{-3}$) or higher B and/or lower P it is the charged pion
radiation which prompts the energy saturation of the protons (this is consistent with the conclusions obtained from the comparison
of the characteristic timescales).

\section{Curvature Pion Neutrino Luminosity}

\label{sec:cpinu}

In terms of the Goldreich-Julian charged particle density
$n_{GJ}=(B/ecP) = 7\times 10^{13}B_{15}P^{-1} \hbox{cm}^{-3}$ for a
pulsar or magnetar of radius $a$ and period $P$ s, with a polar cap
solid angle $\Omega_p=\pi\theta_p^2 \sim (2\pi^2 a/cP)=6.5\times
10^{-4} \a6 P^{-1}$ sr, the outflow rate of ions with charge $Z$ is
\beq 
{\dot N_i}= a^2 \Omega_p n_{GJ} Z^{-1} c \simeq 1.4\times 10^{33} \B15 \a6^3 Z^{-1} P^{-2}~~{\rm s}^{-1}~. \label{eq:dotN}
\enq
For a $\pi^+$ radiation efficiency $\eta_\pi$ at the ion terminal Lorentz factor, 
the resulting number of pion-decay $\nu_\mu$ and subsequent muon-decay 
${\bar \nu}_\mu$ is 2 per proton, giving for a pulsar at distance $D=10$ Kpc a 
$\nu_\mu$ or ${\bar \nu}_\mu$ flux at Earth of
\beq 
\Phi_{\nu,c}= Z {\dot N_i}/ 4\pi D^2 \simeq 1.2 \times 10^{-13} \eta_\pi \B15 P^{-2} \a6^3 D_{10}^{-2}~\hbox{cm}^{-2}\hbox{s}^{-1}, \label{eq:Phinu}
\enq
where the label $c$ denotes the curvature pion origin.  The corresponding 
arrival rate in a detector of area $A=1 A_0$ km$^2$ is ${\dot N_\nu}=\Phi_\nu A 
\simeq 3.7\times 10^4 \eta_\pi \B15 \a6^3 P^{-2} A_0 D_{10}^{-2}$ km$^{-2}$ yr$^{-1}$.

The typical energy of the muon neutrinos is determined by the energy
losses incurred by the pions and muons before the decay. As the pions
move out to radii comparable to the light cylinder, the $\pi^+$
Klein-Nishina energy losses are less important than the $\pi^+$ photon
curvature losses, which occur on a timescale 
$t_{c,\pi}\sim 1.2\times 10^{-12} \eps_{9\pi}^{-3}\a6\P3$ s, where
$\eps_{9\pi}=(\eps_\pi/10^9\hbox{GeV})$ is the pion energy normalized
to $10^9$ GeV, while the $\pi^+$ decay time is 
$t_{d,\pi}\sim 1.86\times 10^2 \eps_{9\pi}$ s.  The pions cool by curvature 
radiation until reaching the light cylinder at $t_L\sim 1.7 \times 10^{-4}\P3$s, 
where their energy has dropped to $\eps_{9\pi L}\sim 1.9\times10^{-3}\a6^{1/3}$. 
Thereafter the pions move in the wind zone in essentially ballistic paths, 
cooling adiabatically on a timescale $t_{ad,\pi}= t_L (\eps_{\pi L} /\eps_{\pi})^{3/2}= 
1.4\times 10^{-8}\P3\a6^{1/2}\eps_{9\pi}^{-3/2}$ s.  The pions decay when their 
energy reaches $\eps_{9\pi d}\simeq 0.89\times 10^{-4} \P3^{2/5}\a6^{1/5}\sim 
10^5$ GeV.  The corresponding $\nu_\mu$ has an energy $\eps_{\nu_\mu} \sim 
\eps_{9,\pi,d}/3 \sim 33 \P3^{2/5}\a6^{1/5}$ TeV. The associated $\mu^+$ starts 
with (2/3) of the decaying pion energy and is subject to adiabatic losses. It 
decays on a timescale $t_{\mu d}\sim 1.97\times 10^4\eps_{9\mu}$ s, where 
$\eps_{9\mu}$ is the muon energy normalized to $10^9$ GeV, reaching at decay 
an energy $\eps_{9 \mu d}\simeq 1.08\times 10^{-5}\P3^{2/5}\a6^{8/25} \sim 10^4$ 
GeV. The associated muon neutrino has an energy $\eps_{\nu_\mu}\sim 
(\eps_{9,\mu,d}/3)\sim 3.3$ TeV.

The neutrino event rate in the detector, however, depends on the mixing of 
neutrino flavors because of oscillations while traveling from their source to Earth. 
Since the muon neutrinos from pion and muon decays are in two different energy 
regimes (33 and 3.3 TeV, respectively), one may consider their flavor oscillations 
independently. With a production flux ratio of $\nu_e$:$\nu_\mu$:$\nu_\tau$=0:1:0, 
for pion decay $\nu_\mu$, the observed ratio would be 0.2:0.4:0.4. For 3.3 TeV muon 
decay neutrino fluxes, with a production ratio of 1:0:0 and 0:1:0, respectively 
for $\nu_e$ and ${\bar \nu}_\mu$, the respective observed ratios would be 
0.6:0.2:0.2 and 0.2:0.4:0.4.  Since the neutrino detectors cannot distinguish 
between a muon neutrino and an anti-neutrino, the observed flux of 3.3 TeV 
muon neutrino would be a factor 0.6 times the production flux. The probability 
per $\nu_\mu$ of resulting in an upward muon is $P_{\nu\to\mu} \sim 1.3 \times 
10^{-6} (\eps_{\nu_\mu} /\hbox{TeV})$, which for the 33 and 3.3 TeV neutrino 
cases lead to upward muon count rates from curvature pions, after taking into 
account oscillation effects, of
\beq 
{\dot N}_{\mu ,c} \simeq \begin{cases} 6.3 \times 10^5 \eta_\pi \B15 
  \P3^{-8/5}\a6^{10/3} A_0 D_{10}^{-2} (\eps_{\nu_\mu}/33\hbox{TeV})~ 
  \hbox{km}^{-2} \hbox{yr}^{-1}, \cr
9.5 \times 10^4 \eta_\pi \B15 \P3^{-8/5}\a6^{10/3} A_0 D_{10}^{-2} 
  (\eps_{\nu_\mu}/3.3\hbox{TeV})~ \hbox{km}^{-2} \hbox{yr}^{-1},
\end{cases}
\label{eq:mupic}
\enq
for an on-beam magnetar at 10 Kpc.

In the case of proton acceleration, from Fig.  \ref{fig:tp} these
would reach a maximum energy determined by $\pi^+$ curvature losses
for fields $B\simg 10^{15}$ G and $P\sim 10^{-3}$ s, or
$\B15\P3^{-2}\simg 1$. In these cases $t_{\pi^+} \siml t_c$ and the
$\pi^+$ radiation efficiency $\eta_\pi\sim 1$, so the upward muon
event rate should be detectable in the 1-100 TeV range with cubic
kilometer detectors, if the pulsars is on-beam and 
$\B15 \P3^{-8/5} A_0 D_{10}^{-2} \simg 10^{-4}$, e.g. for sources in 
the local group. On the other hand, for values $\B15 \P3^{-2} < 1$, 
the pion efficiency is low, $\eta_\pi \sim t_c/t_\pi < 1$, which
would allow potentially detectable neutrino fluxes for sources in our
own galaxy provided $\eta_\pi \simg 10^{-4}-10^{-5}$. For
$\B15\P3^{-2}\ll 10^{-5}$, the steep behavior of the pion radiation
efficiency implies an unobservably small number of events, even for
galactic sources. Thus, there is a small range of magnetar field
strengths $\B15\simg 1$ and (short) periods $\P3\sim 1$ for which
these objects could be potentially detectable neutrino targets.


\section{Photomeson Neutrinos, Cosmic Rays and Photons}

\label{sec:photomes}

Another mechanism for neutrino production in the polar caps is the
photomeson process $p\gamma \to \pi^+ \to \mu^+\nu_\mu \to e^+\nu_e{\bar \nu}_\mu$, 
acting on the magnetar surface X-rays, typically $L_x\sim 10^{35} \L35$ 
erg s$^{-1}$ (Zhang et al, 2003). The photon number density near the surface 
is $n_{\gamma,a}= 1.8\times 10^{21} \L35^{3/4}$ cm$^{-3}$, and for protons 
of energy much above the threshold $\eps \gg 0.1-0.2$ GeV, the cross section
$\sigma_{p\gamma}\sim 10^{-28}$ cm$^2$ implies a mean interaction time
$t_{p\gamma}\simeq 1.8\times 10^{-4}\L35^{-3/4}$ s, a mean free path
$\ell_{p\gamma}\sim 5\times 10^6 \L35^{-3/4}$ cm, or an optical depth
$\tau_{p\gamma}\sim 0.18\L35^{3/4}\a6$.  For $\L35\simg 1$ and any but
the fastest rotating pulsars, the photomeson interaction occurs on a
length comparable or less than the light cylinder $R_L=5\times 10^6 \P3$ cm, 
but much larger than the height $h$ where acceleration balances pion or 
photon curvature losses.

If the pion curvature process is inefficient, $\eta_\pi\ll 1$, most of
the protons survive beyond $h$, and inside the light cylinder they
cool due to curvature radiation (or outside the light cylinder due to
adiabatic cooling), until the photomeson process leads to a neutron
and a $\pi^+$ which takes $0.2$ of the remaining proton energy. The
pions and the subsequent muons then cool due to curvature (if inside
$R_L$) and adiabatic (when outside $R_L$) losses in a similar way as
in the previous section. For typical parameters, the average neutrino
energy is $<\eps_{\nu_\mu}>\simeq 28 \P3^{3/5}\L35^{-3/20} \a6^{1/5}$
TeV, not much different from the value $\sim 18 \P3^{2/5}\a6^{8/25}$
TeV in the pion curvature initiated case of last section, since both
the protons (before undergoing $p\gamma$) and the charged pions cool
by photon curvature radiation inside the light cylinder, and
adiabatically outside.  The neutrino flux resulting from $p\gamma$ are
larger, for $\tau_{p\gamma} \sim 0.2 \gg \eta_\pi$, and since the
typical neutrino energies from $p\gamma$ are typically only $\sim 2$
larger than in the pion curvature case, in this $\eta_\pi \ll 1$ limit
it would be hard to distinguish the pion curvature component from the
dominant $p\gamma$ flux.

The situation is different for large $\eta_\pi \to 1$. In this case it
is mainly neutrons that propagate outwards beyond $h$. Being neutral,
they do not suffer adiabatic losses, and undergo photopion
interactions $n\gamma \to \pi^-$ on a timescale similar to protons,
$t_{n\gamma}\simeq 1.8\times 10^{-4}\L35^{-3/4}$ s.  Typically this
occurs inside or at the light cylinder. We take as a limiting example
the case where this occurs near the light cylinder. The negative pions
and muons then undergo adiabatic cooling outside $R_L$ similarly to
their positive counterparts, as discussed previously, resulting in
${\bar \nu}_\mu$. Ignoring the distinction between ${\bar \nu}_\mu$
and $\nu_\mu$, which cannot be discriminated in Cherenkov detectors,
the resulting average neutrino energy is now $<\eps_{\nu_\mu}>\sim 
280 \L35^{-3/10}\eps_{9ni}^{3/5}$ TeV, where
$\eps_{9ni}=(\eps_n/10^9\hbox{GeV})$ is the initial neutron energy. 
The flux is $\Phi_{\nu n\gamma}\sim 5\times 10^{-8} \P3^{-2}\B15 \a6^4 
\L35^{3/4} \D10^{-2}~\hbox{cm}^{-2}\hbox{s}^{-1}$ around $\eps_\nu\sim 280$ 
TeV from $n\gamma$ interactions, and the upward muon event rate, 
 modulo the factors due to oscillation effect as previously discussed, is
\beq 
{\dot N}_{\mu ,n\gamma} \simeq 5.4 \times 10^6 \B15 \P3^{-2}\a6^{4} 
\L35^{9/20} A_0 D_{10}^{-2} (\eps_{\nu_\mu}/280\hbox{TeV})~ 
\hbox{km}^{-2} \hbox{yr}^{-1}, 
\label{eq:mungam}
\enq
for an on-beam magnetar at 10 Kpc, from $n\gamma$ interactions. The
corresponding pion curvature neutrinos would lead to an upward muon
event rate given by equation (\ref{eq:mupic}), which is not too
different. If observed, the difference in the average neutrino 
energies from pion curvature ($\sim 18$ TeV) and from $n\gamma$ 
interactions ($\sim 280$ TeV) could help discriminate between the
pion curvature and the photomeson production mechanisms. If there 
were such milisecond magnetars with $\eta_\pi\to 1$ in our galaxy 
($D \siml 10$ Kpc) this would imply, from eqs. (\ref{eq:mupic},
\ref{eq:mungam}) upward muons event rates which would be strong
in a completed cubic kilometer detector, and probably even in
partially completed installations, so the current presence of 
such objects in our galaxy  is questionable.

These ${\vec{\Omega}\cdot \vec{B}} <0$ caps will also result in a
significant flux of escaping neutrons, which contribute to the cosmic
ray flux.  As an example, taking the case for the efficiency of
$\pi^+$ radiation by protons from magnetars with $\B15\P3^{-2}=1$, for
which $\eta_\pi\simeq t_{\pi^+}/t_c\sim 0.2$, the flux of neutrons at
Earth and their energies are
\beq 
\Phi_n \simeq 10^{-11} \B15 \P3^{-2} D_{10}^{-2} ~{\rm m}^{-2}~{\rm s}^{-1}~
{\rm sr}^{-1} ~~;~~ \epsilon_n \simeq 6\times 10^{17}~{\rm  eV} ~.  
\enq
A neutron of $6\times 10^{17}$~eV energy decays after traveling $\sim 6$~kpc.  
 For comparison, the observed cosmic-ray number flux at this energy is 
$\approx 9\times 10^{-12}~{\rm m}^{-2}~{\rm s}^{-1}~{\rm sr}^{-1}$ 
(Nagano \& Watson, 2000). However, the very small degree of observed cosmic 
ray anisotropy at these energies again suggests that the probability of 
finding such $\eta_\pi\to 1$ magnetars within our galaxy at any given 
time is very low.

In the case of a newly born neutron star or magnetar, the escaping
neutrons may interact with the expanding supernova-remnant (SNR)
shell, as discussed by Razzaque, \Mesz \& Waxman (2003).  For a
hypernova which results in a magnetar, the shell speed is $v = 
10^{10}v_{10}$~cm/s and reaches a radius $R_{snr} \simeq 8.6\times 
10^{14} v_{10} t_d$~cm in $t_d$ days.  With a typical shell mass of
$M_{snr} = 2\times 10^{33} m_{snr}$~g, the column density of target
atoms is $\Sigma_A \simeq 1.3\times 10^{26} m_{snr} t_d^{-2} v_{10}^{-2}~
{\rm cm}^{-2}$.  The $pp$ cross-section for $\sim 10^{18}$~eV energy 
incident neutrons is 100~mb and the corresponding
optical depth is $\tau_{pp}\sim 10 m_{snr} t_d^{-2} v_{10}^{-2}$.  The
energy of the neutrinos produced by a $7\times 10^{17}$~eV neutron via
$pp$ interactions ranges from $\sim m_\pi c^2 \gamma_{cm}/4 \simeq 0.7$~TeV 
up to $\sim \eps_n/4$ with a $\eps_\nu^{-1}$ distribution normalized to 
a multiplicity of $10^{-3}$ at this energy (Razzaque,
\Mesz \& Waxman, 2003).  Here $\gamma_{cm}$ is the Lorentz factor of
the center-of-mass of the $pp$ interaction in the Lab frame.

There may also be photo-hadronic interactions with photons in the SNR
shell created from the SN explosion.  The peak energy of these
blackbody photons at creation is $\approx 18.2 E_{51}^{1/4} r_{11}^{3/4}$~keV 
for $\sim 10^{51}E_{51}$~erg SN explosion energy and $\sim 10^{11} r_{11}$~cm 
progenitor star's radius. In the SNR shell, however, they cool down to a 
peak energy of $\eps_{\gamma,sn} \simeq 2.1 E_{51}^{1/4} r_{11}^{1/4} 
v_{10}^{-1} t_d^{-1}$~eV.  Thus neutrons of $\gtrsim 10^{17}$~eV energy 
may produce pions by interacting with them.  Assuming that the total number 
of SN photons do not change after their creation, the column density of 
them in the SNR shell is $\Sigma_{\gamma,sn} \simeq 3.65\times 10^{27} 
E_{51}^{3/4} r_{11}^{3/4} v_{10}^{-2} t_d^{-2}~{\rm cm}^2$ and the opacity for
photo-hadronic interactions is $\lesssim 1$ with 0.1~mb typical
cross-sections at the resonances.  The resulting neutrino energy in
this case would be in the 10's of PeV range. There may be additional
components of VHE ions accelerated in the SNR shell or the outer-gaps
of a pulsar/magnetar which could also interact with the SNR shell
(Guetta \& Granot, 2003). These interactions will produce additional
neutrinos and hence identifying the neutrinos from pion curvature
radiation may be difficult for $t\siml$ few days after a magnetar is
born.

Ultra-high energy photons will also be produced, both through
curvature radiation of charged pions and muons, and through decay of
the associated neutral pions. E.g. pions produced by $\sim 10^9$ GeV
protons would lead to pion-related UHE photons of luminosity
$L_{\gamma (\pi)}\sim 10^{44} \B15 \P3^{-2} \a6^3$ erg s$^{-1}$,
decaying as the field drops and the period lengthens. The curvature
photon energy is $\eps_\gamma \sim 3 \P3^{-2} \a6^{-1/2} (\eps_{\pi}/
10^8\hbox{GeV})^3$ PeV. These, as well as the muon curvature and neutral 
pion decay photons are all well above the $2m_e c^2/ \sin\theta$ 
threshold for one-photon pair production $\gamma B \to e^\pm$ (e.g. 
Harding and Lai, 2006), where $\theta$ is the angle of propagation 
relative to field direction.  The optical depth above threshold for 
a path length $10^6R_6$ cm is $\tau_{\gamma B}\sim 5\times 10^{10} 
(\B15^{2/3}\sin\theta)^{2/3} R_6 (\eps_\gamma /\hbox{PeV})^{-1/3}$. 
(Another UHE photon opacity is photon splitting,
a higher order mechanism which above threshold is less important than
one-photon pair formation).  The UHE photons will thus be degraded to
energies below the threshold, $\eps_\gamma\siml 10~ \theta_{-1}^{-1}$
MeV, for cap angles $\theta \sim 10^{-1}\theta_{-1}$.

\section{Discussion}
\label{sec:disc}

Our discussion has assumed that the properties of the inner gaps near
the polar caps of pulsars or magnetars allow protons to be accelerated
to energies $\simg 6\times 10^8$ GeV. This is an open question, since
the extent and properties of polar gaps (with ${\vec \Omega}\cdot 
{\vec B}\leq 0$) including self-consistently pion radiation effects is 
unknown. Such studies, if undertaken, would have to take into account
not only the effects of leptons acceleration from charged pion decays,
but also high energy photons from neutral pion decay and their 
subsequent cascades.  Studies of space charge limited gaps with
electron acceleration and pair production (e.g.  Harding and Muslimov,
2002; also Baring and Harding, 2002) suggest that Lorentz factors of
order $10^9$ may be possible in fast-rotating, high field objects, in
which case the effects discussed here may become important.

\begin{figure*}

\plotone{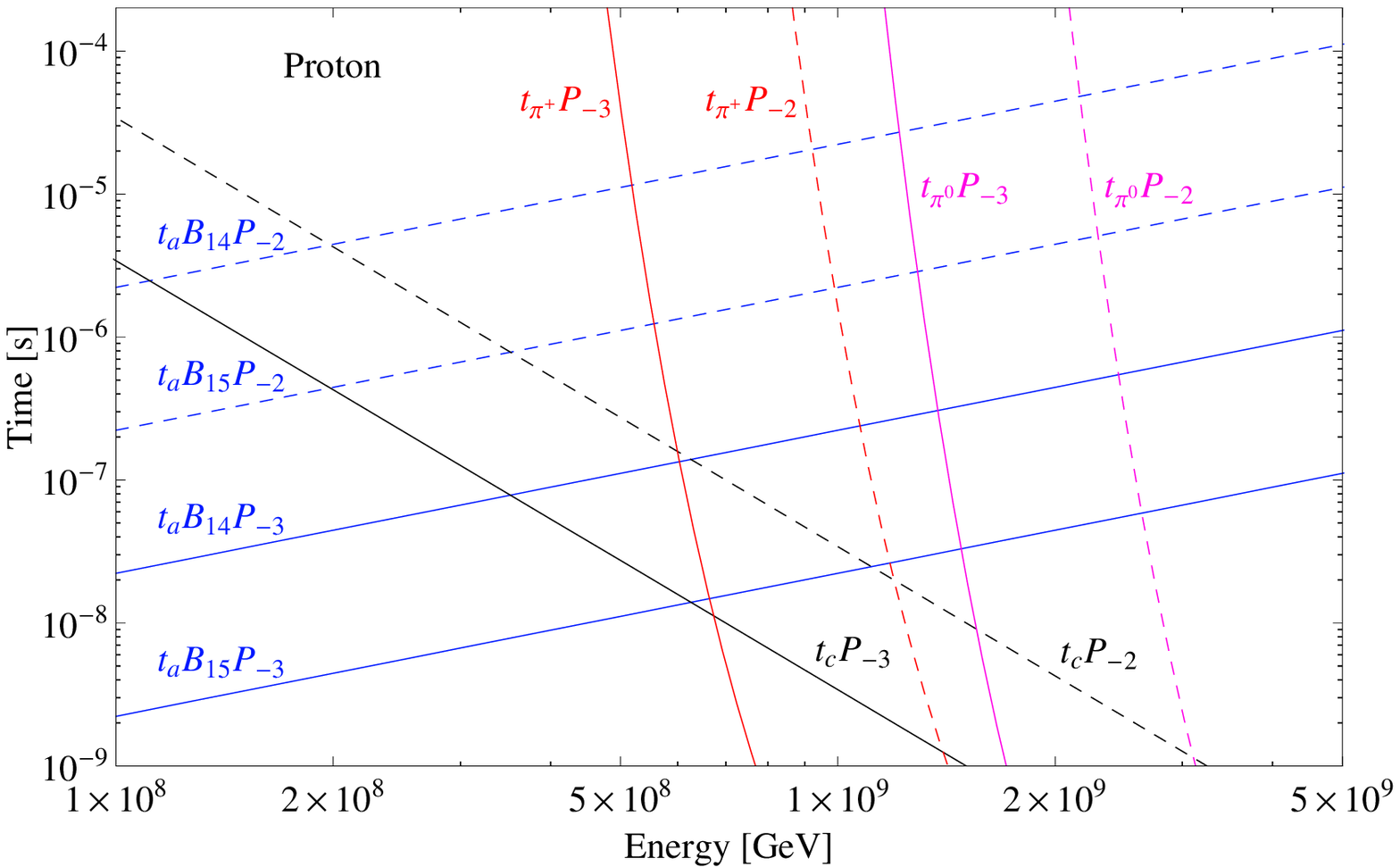} 

\caption{ Energy loss timescales $t_{\pi^+}$ and $t_{\pi^0}$ for proton
curvature emission of $\pi^+$ and $\pi^0$, compared to photon
curvature radiation loss timescale $t_c$ and acceleration time $t_a$,
as a function of proton energy (in GeV), for different field strengths
$B=10^{15}\B15$ (G) and periods $P=10^{-3}\P3$ (s).  }
\label{fig:tp}
\end{figure*}

The curvature pion radiation mechanism is likely to be of interest for
fast rotating magnetars which accelerate protons. This is because,
from Fig.~\ref{fig:tp}, we see that for magnetic field and period
values $\B15 \P3^{-1} \simg 1$ the pion curvature radiation dominates
the photon curvature radiative cooling.  
 For such magnetic field and period combinations, the acceleration 
of heavy ions results in the fragmentation of the ions before pion
emission occurs, which is expected only from the individual protons 
resulting from the fragmentation after they have been further accelerated.

There is so far no observational evidence for millisecond  periods
among the handful of known magnetars (e.g. Kaspi, 2007; Woods and Thompson, 
2004), most of which are in our galaxy and have periods of seconds.  
From eq. (\ref{eq:mupic}), we see that for $B\simg 10^{15}$~G and 
millisecond periods, the muon event rate is so large that  they  
might have been detected by now, if the  magnetar were on-beam
and at a distance $D\siml$ Mpc.  The likeliest explanation is that
such millisecond magnetars do not exist in our galaxy (or else they
might be off-beam). At the same time, it is reassuring that  for
periods $P\simg 1$ s  the pion curvature efficiency predicted is
$\eta_\pi=t_c/t_\pi \lll 1$, and the corresponding muon event rate 
in cubic kilometer detectors given by equation (\ref{eq:mupic}) 
is negligible for known galactic magnetars.

 An interesting possibility is that some fraction of core collapse 
supernovae lead, at least initially, to a millisecond magnetar. There 
is currently no observational evidence for this, but it is a 
plausible  hypothesis, among others from fast convective overturn 
dynamo arguments, e.g. Thompson and Duncan (1996).  In such objects, 
the envelope optical thickness to $\nu N$ interactions is $\tau_{\nu  N}
\sim 10^{-6} (\eps_\nu /\hbox{10 TeV}) (M_{env}/10\msun) (v_{env}/0.1c)^{-2} 
(t/\hbox{day})^{-2}$, and as long as the magnetic field and the 
rotation rate remain high, neutrinos produced by pion curvature and 
$p\gamma$ interactions can escape without further reprocessing.  
 Thus, core collapse supernovae which result initially in a millisecond
magnetar with $B\simg 10^{15}$ G are expected to undergo a brief period
of intense pion and ultra-high energy neutrino production, which
significantly exceeds any electromagnetic energy losses. If the
fraction of core collapse supernovae leading to such objects were
$0.05-0.1$, given the frequency of core collapse SNe in the
LMC and in the local group, one could in principle expect some
millisecond magnetars detectable in cubic kilometer detectors within
timescales of years.  In this scenario one would also expect (\S
\ref{sec:photomes}) a significant $\siml 10$ MeV photon luminosity,
which may be detectable by the GLAST GBM.

In summary, we have pointed out that pions produced by protons 
interacting with the curved strong magnetic fields of magnetars may 
be an important energy loss mechanism for the accelerated particles, 
leading to secondary leptons and photons which can affect the 
properties of the inner gaps.  Our model for the strong
interaction processes involved here relies on the softness of the pion
radiation process. Although the correctness of this assumption needs
further, more detailed investigation, this level of approximation is 
justifiable here, given the larger uncertainties in the astrophysical 
model. Our results indicate that the effects analysed in this paper
could lead to copious neutrino production in the TeV-nPeV range, which 
would provide interesting targets for cubic kilometer scale detectors.

{\acknowledgments This work is supported in part by NSF AST 0307376
and the Hungarian Science Foundation OTKA No. T046129 and T68108. SR is presently
a National Research Council Research Associate at the Naval Research
Laboratory.  We are grateful to A.K.  Harding and to the referee for 
useful comments.}

\end{document}